\documentclass[twoside]{article}

\usepackage{graphics}
\usepackage[sort]{natbib}
\usepackage{amsmath}
\usepackage{floatflt}
\usepackage{rotating}

\renewcommand{\sectionmark}[1]{\markboth{\textsc{M. E. J. Newman and
M. Girvan}}{\textsc{Mixing patterns and community structure in networks}}}
\renewcommand{\subsectionmark}[1]{\markboth{\textsc{M. E. J. Newman and
M. Girvan}}{\textsc{Mixing patterns and community structure in networks}}}

\newcommand{\captionfonts}{\small}
\makeatletter
\long\def\@makecaption#1#2{%
  \vskip\abovecaptionskip
  \sbox\@tempboxa{{\captionfonts #1: #2}}%
  \ifdim \wd\@tempboxa >\hsize
    {\captionfonts #1: #2\par}
  \else
    \hbox to\hsize{\hfil\box\@tempboxa\hfil}%
  \fi
  \vskip\belowcaptionskip}
\makeatother

\bibpunct{[}{]}{,}{n}{}{,}

\newcommand{\half}{\mbox{$\frac12$}}

\newcommand{\ord}{{\rm O}}
\newcommand{\e}{{\rm e}}

\newcommand{\av}[1]{\langle#1\rangle}
\newcommand{\eref}[1]{(\ref{#1})}
\newcommand{\etal}{{\it{}et~al.}}
\newcommand{\ve}{\mathbf{e}}
\newcommand{\vx}{\mathbf{x}}
\newcommand{\cN}{\mathcal{N}}
\newcommand{\Tr}{\mathop{\rm Tr}}
\newcommand{\norm}[1]{||\,#1\,||}
\newcommand{\phm}{\phantom{-}}

\begin{document}

\title{Mixing patterns and community structure\\
in networks}
\author{M. E. J. Newman$^{1,3}$ and M. Girvan$^{2,3}$\\
\\
\textit{\small $^1$Department of Physics, University of Michigan,
Ann Arbor, MI 48109.  U.S.A.}\\
\textit{\small $^2$Department of Physics, Cornell University, Ithaca,
NY 14853.  U.S.A.}\\
\textit{\small $^3$Santa Fe Institute, 1399 Hyde Park Road, Santa Fe,
NM 87501.  U.S.A.}}
\date{}
\maketitle

\begin{abstract}
  Common experience suggests that many networks might possess community
  structure---division of vertices into groups, with a higher density of
  edges within groups than between them.  Here we describe a new computer
  algorithm that detects structure of this kind.  We apply the algorithm to
  a number of real-world networks and show that they do indeed possess
  non-trivial community structure.  We suggest a possible explanation for
  this structure in the mechanism of assortative mixing, which is the
  preferential association of network vertices with others that are like
  them in some way.  We show by simulation that this mechanism can indeed
  account for community structure.  We also look in detail at one
  particular example of assortative mixing, namely mixing by vertex degree,
  in which vertices with similar degree prefer to be connected to one
  another.  We propose a measure for mixing of this type which we apply to
  a variety of networks, and also discuss the implications for network
  structure and the formation of a giant component in assortatively mixed
  networks.
\end{abstract}

\thispagestyle{empty}

\section{Introduction}
Much of the recent research on the structure of networks of various kinds
has looked at properties like path lengths, transitivity, degree
distributions, and resilience of networks to vertex
deletion~\cite{Strogatz01,AB02,DM02}, all of which, while of exceptional
importance in many contexts, tend to focus our attention on the properties
of individual vertices or vertex pairs---how far apart they are, what their
degrees are, and so forth.  However, in other contexts it may be equally
important to ask about the large-scale properties of the network as a
whole.  Numbers of components and their distribution of sizes would be an
example of such a property, one which is relevant to issues of
accessibility~\cite{Broder00} and to
epidemiology~\cite{Grassberger83,BM90,MPV02}.  Searchability and the
performance of search algorithms on networks would be
another~\cite{Kleinberg00proc,ALPH01,WDN02}.  A third is the existence and
effects of large-scale inhomogeneity in networks---what we call ``community
structure'', the presence (or absence) in the network of regions with high
densities of connections between vertices and other regions with low
densities---and it is with a discussion of this topic that we begin this
paper.  (In some circles, this phenomenon is called ``clustering'', an
unfortunate terminology which risks confusion with another use of the word
clustering introduced recently by Watts and Strogatz~\cite{WS98}.  We will
use the word clustering only in reference to hierarchical clustering, which
is a standard technique for community detection; otherwise we will avoid
it.)  Our investigation of community structure will lead us to
consideration of mixing patterns in networks---which vertices connect to
which others and why---as an explanation for observed communities in
networks of all kinds, and eventually to consideration of more general
classes of correlated networks including networks with correlations between
the degrees of adjacent vertices.

Much of the work reported in this article has appeared previously in
various papers, which the reader may like to consult for more detail than
we can give here~\cite{GN02,Newman02f,Newman02g}.

\section{Community structure}
\label{community}
The oldest studies by far of the large-scale statistical properties of
networks are the studies of social networks carried out within the
sociological community, which stretch back at least to the
1930s~\cite{WF94,Scott00}.  Social networks are network representations of
relationships of some kind, generically called ``ties'', between people or
groups of people, generically called ``actors''.  Actors might be
individuals, organizations or companies, while ties might represent
friendship, acquaintance, business relationships or financial transactions,
amongst other things.

\begin{floatingfigure}{5.2cm}
\begin{center}
\resizebox{4.7cm}{!}{\includegraphics{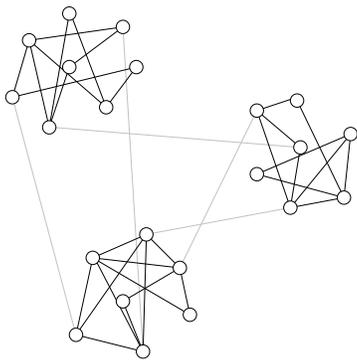}}
\end{center}
\caption{A figurative sketch of a network possessing community structure of
the type discussed here.}
\label{sketch}
\end{floatingfigure}

A long-standing goal among social network analysts has been to find ways of
analysing network data to reveal the structure of the underlying
communities that they represent.  It is commonly supposed that the actors
in most social networks divide themselves naturally into groups of some
kind, such that the density of ties within groups is higher than the
density of ties between them.  A sketch of a network with such community
structure is shown in Fig.~\ref{sketch}.

It is a matter of common experience that social networks do contain
communities.  We look around ourselves and see that we belong to this
clique or that, that we have a circle of close friends and others whom we
know less well, that there are groupings within our personal networks on
the basis of interest, occupation, geographical location and so forth.
This does not however guarantee that a network contains community structure
of type that we are considering here.  It would be perfectly possible for
each person in a network to have a well-defined set of close acquaintances,
their own personal network neighbourhood, but for the network
neighbourhoods of different people to overlap only partially, so that the
network as a whole is quite homogeneous, with no clear communities emerging
from the pattern of vertices and edges.  A network model showing precisely
this type of structure has been proposed and studied recently by
Kleinberg~\cite{Kleinberg02}.  Our purpose in this section will be to
investigate methods for detecting whether true community structure does
exist in networks and for extracting the communities, and to apply those
methods to particular networks.  As we will see, the early intuition of the
sociologists was correct, and many of the networks studied, including
non-social networks, do possess large-scale inhomogeneity of precisely the
type that would indicate the presence of community divisions.

The problem then is to take a network, specified in the simplest case by a
list of $n$ vertices joined in pairs by $m$ edges, and from this structure
to extract a set of communities---non-overlapping subsets of vertices that
are, in some sense, tightly knit, having stronger within-group connections
than between-group connections.  The traditional, and still most common,
method for detecting structure of this kind is the method of ``hierarchical
clustering''~\cite{WF94,Scott00}.  In this method one defines a connection
strength for each pair of vertices in the network, i.e.,~$\half n(n-1)$
numbers that represent a distance or weight for the connection between each
pair.  (In some versions of the method not all pairs are assigned a
connection strength, in which case those that are not can be assumed to
have a connection strength of zero.)  Examples of possible definitions for
the strengths include geodesic (shortest path) distances between pairs, or
their inverses if one wants a measure that increases when pairs are more
closely connected, counts of numbers of vertex- or edge-independent paths
between pairs (``maxflow'' methods) or weighted counts of total numbers of
paths between pairs (adjacency matrix methods).

\begin{floatingfigure}{6.3cm}
\begin{center}
\resizebox{5.8cm}{!}{\includegraphics{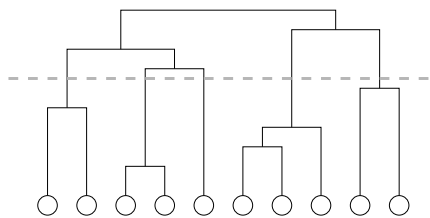}}
\end{center}
\caption{An example of a dendrogram showing the hierarchical clustering of
ten vertices.  A horizontal cut through the dendrogram, such as that
denoted by the dotted line, splits the vertices into a set of communities,
five in this case.}
\label{dendrogram}
\end{floatingfigure}

Then, starting with the $n$ vertices but no edges between them, one joins
vertices together in decreasing order of the weights of vertex pairs,
ignoring the edges of the original network.  One can pause at any stage in
this process and observe the pattern of components formed by the
connections added so far, which are taken to be the communities of the
network at that stage.  The heirarchical clustering method thus defines not
just a single decomposition of the network into communities, but a nested
hierarchy of possible decompositions, having varying numbers of
communities.  This hierarchy can be represented as a tree or
``dendrogram'', an example of which is shown in Fig.~\ref{dendrogram}.  A
horizontal cut through the dendrogram at any given height, such as that
denoted by the dotted line in Fig.~\ref{dendrogram}, splits the tree into
the communities for the corresponding stage in the hierarchical clustering
process.  By varying the height of the cut, one can arrange for the number
of communities to take any desired value.

The construction of dendrograms is a popular technique for the analysis of
network data, particularly within the sociological community.  Software
packages for network analysis, such as Pajek and UCInet, incorporate
hierarchical clustering as a standard feature: for any network one can
calculate a huge variety of vertex--vertex weights of different types and
construct the corresponding dendrogram for any of them.  The method however
has some problems.  There are many cases in which networks have rather
obvious community structure, but hierarchical clustering fails to find it.
One particular pathology that is frequently observed is that peripheral
vertices tend to get disconnected from the bulk of the network, rather than
being associated with the groups or communities that they are primarily
attached to.  For example, if a vertex is connected to the rest of the
network by only a single edge, then presumably, were one to assign it to a
community, it would be assigned to the community that the single edge leads
to.  In many cases, however, the hierarchical clustering method will
declare the vertex instead to be a single-vertex community in its own
right, in disagreement with our intuitive ideas of community structure.

In a recent paper therefore~\cite{GN02} we have proposed an alternative
method for detecting community structure, based on calculations of
so-called edge betweenness for vertex pairs.  As we will see, this method
detects the known community structure in a number of networks with
remarkable accuracy.

\subsection{Edge betweenness and community detection}
\label{gn}
Freeman~\cite{Freeman77} proposed a measure of centrality for the actors in
a social network which he called ``betweenness''.  The betweenness of an
actor is defined to be the number of shortest paths between pairs of
vertices that pass through that actor.  In cases where the number $p$ of
shortest paths between a vertex pair is greater than one, each path is
given an equal weight of~$1/p$.  Trivial algorithms for calculating
betweenness take $\ord(mn^2)$ time to calculate betweenness for all vertices,
or $\ord(n^3)$ time on a sparse graph (i.e.,~one in which the number of edges
per vertex is constant in the limit of large graph size).  This makes the
calculation prohibitively costly on large networks.  Recently however, two
new algorithms have been proposed~\cite{Newman01c,Brandes01} that both
allow the same calculation to be performed faster, in time $\ord(mn)$, or
$\ord(n^2)$ on a sparse graph, by eliminating needless recalculations of
geodesic paths.  The betweenness of a vertex gives an indication, as the
name implies, of how much the vertex is ``between'' other vertices.  If,
for example, information (or anything else) spreads through a network
primarily by following shortest paths, then betweenness scores will
indicate through which vertices most information will flow on average.  The
vertices with highest betweenness are also those whose removal will result
in an increase to the geodesic distance between the largest number of other
vertex pairs.

Here we consider an extension of Freeman's betweenness to the edges in a
network.  The betweenness of an edge is defined to be the number of
shortest paths between pairs of vertices that run along that edge, with
paths again being given weights $1/p$ when there are $p>1$ between a given
pair of vertices.  In fact, the concept of edge betweenness actually
appears to predate Freeman's work on vertex betweenness, having appeared in
an obscure technical report by an Amsterdam mathematician some years
earlier~\cite{Anthonisse71}.  Edge betweenness has received very little
attention in other literature until recently, but it provides us with an
excellent measure of which edges in a network lie between different
communities.  In a network with strong community structure---groups of
vertices with only a few inter-group edges joining them---at least some of
the inter-group edges will necessarily receive high edge betweenness
scores, since they must carry the geodesic paths between vertex pairs that
lie in different communities.  This implies that eliminating edges with
high edge betweenness from a graph will remove the inter-group edges, and
hence split the graph efficiently into its different groups.  This is the
principle behind our method for the detection of community structure.  Our
algorithm is as follows.
\begin{enumerate}
\item We calculate the edge betweenness of every edge in the network.
\item We remove the edge with the highest betweenness score, or randomly
choose one such if more than one edge ties for the honour.
\item We recalculate betweenness scores on the resulting network and repeat
from step~2 until no edges remain.
\end{enumerate}
The recalculation in step~3 is crucial to the method's success.  When there
is more than one inter-group edge between two groups of vertices, there is
no guarantee that both will receive high betweenness scores; in some cases
most geodesic paths will flow along one edge and only that one will receive
a high score.  Recalculation ensures that at some stage in the working of
the algorithm each inter-group edge receives a high score and thus gets
removed.

The calculation of all edge betweennesses takes time $\ord(mn)$, and its
repetition for all $m$ edges thus gives the algorithm a worst-case running
time of~$\ord(m^2n)$, or $\ord(n^3)$ on a sparse graph.  The results of the
algorithm can be represented as a dendrogram, just as in traditional
hierarchical clustering, although one should be aware that the construction
of the tree is not logically the same: the recalculation of the
betweennesses after each edge removal means that there is no single
function that can be defined for each edge in the initial graph such that
the resulting dendrogram is the representation of a hierarchical clustering
construction carried out using that function.

\subsection{Examples}
\label{examples}
\begin{figure}[t]
\begin{center}
\resizebox{8cm}{!}{\includegraphics{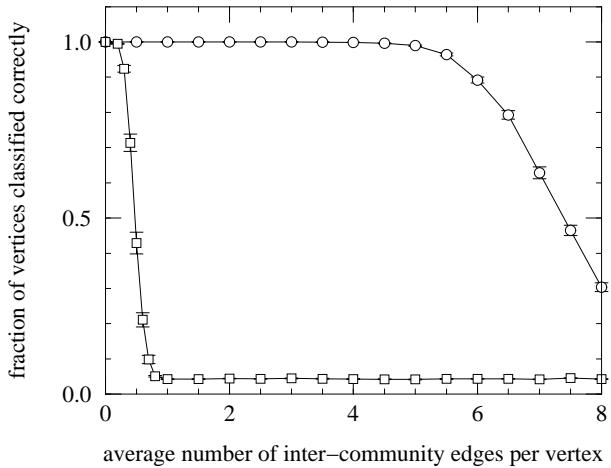}}
\end{center}
\caption{The fraction of vertices correctly classified in applications of
community structure finding algorithms to the computer-generated graphs
described in the text.  The circles are results for the method presented in
this paper and the squares are for the standard hierarchical clustering
method, using a maximum-flow measure of connection strength between vertex
pairs.  Each point is an average over 100 realizations of the graphs.}
\label{randtest}
\end{figure}

Here we give three examples of the application of our community structure
finding algorithm to different networks.  The first example is a set of
computer generated graphs, specifically created to test the algorithm.  We
created a large number of graphs of 128 vertices each, divided into four
groups of 32 vertices.  Edges were placed at random between vertices within
the same group with probability~$p_{\rm in}$ and between vertices in
different groups with probability $p_{\rm out}$, with the values of $p_{\rm
in}$ and $p_{\rm out}$ chosen to make the average degree of a vertex equal
to~16, and $p_{\rm out}\le p_{\rm in}$.  These graphs were then fed into
our community structure algorithm, and we measured what fraction of the
vertices were correctly classified into their communities as a function of
the ratio of $p_{\rm in}$ to $p_{\rm out}$, or equivalently the mean number
$z_{\rm out}$ of edges from a vertex to vertices in other communities.  The
results are shown in Fig.~\ref{randtest}.  As the figure shows, the
algorithm performs almost perfectly for values of $z_{\rm out}$ up to about
6.  Beyond this point, as $z_{\rm out}$ approaches the value of~8 at which
each vertex has as many inter-group edges as intra-group ones, the fraction
of successfully classified vertices falls off sharply.

On the same plot we also show the performance of a standard hierarchical
clustering algorithm based on edge-independent path counts (maxflow) on the
same set of random graphs.  As the figure shows, the traditional method is
far inferior to our new algorithm in finding the known community structure.

\begin{figure}[t]
\begin{center}
\resizebox{9cm}{!}{\includegraphics{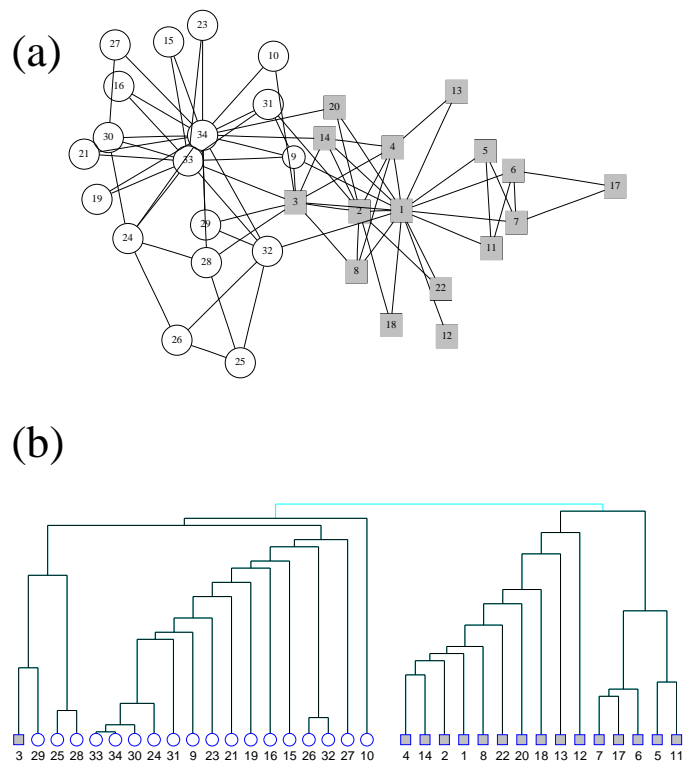}}
\end{center}
\caption{(a)~The friendship network given by Zachary~\cite{Zachary77} for
his karate club study.  Grey squares represent individuals who in the
fission of the club sided with the club's instructor, while open circles
represent individuals who sided with the club's president.  (b)~The
dendrogram representing the community divisions found by our method for
this network, with the same colouring scheme for the vertices.}
\label{zachary}
\end{figure}

For our second example, we move to real-world network data.  In 1977, Wayne
Zachary published the results of an ethnographic study he had conducted of
social interactions between 34 members of a karate club at an American
university~\cite{Zachary77}.  He recorded social contacts between members
of the club over a two year period and published his results in the form of
social networks.  Fortuitously there arose, during the course of the study,
a dispute between the two leaders of the club, the karate teacher and the
club's president, over whether to raise the club's fees.  Ultimately, the
dispute resulted in the departure of the karate teacher and his starting
another club of his own, taking with him about a half of the original
club's members.  Here we analyse a network constructed by Zachary of
friendships between club members before the split occurred.  We compare the
predictions of our community-finding algorithm applied to this network with
the known lines along which the club divided.  Our results are shown in
Fig.~\ref{zachary}.

In panel~(a) of the figure we show the original network, with the grey
squares representing the faction that ultimately sided with the teacher
(who is vertex number~1), and the open circles the faction that sided with
the club's president (vertex number~34).  In panel~(b) we show the
dendrogram output by our algorithm for this network.  As the figure shows,
the algorithm again performs nearly perfectly, with only one vertex, vertex
number~3, being misclassified.  (Inspection of panel~(a) reveals that
vertex~3 is in fact precisely caught in the middle of the network between
the two factions, and so it is not entirely surprising that this vertex was
misclassified.)  Bear in mind that the network in this example was recorded
\emph{before} the fission of the club, so that the results of panel~(b) are
in some sense a prediction of events that were, at that time, yet to occur.

\begin{figure}[t]
\begin{center}
\resizebox{9cm}{!}{\includegraphics{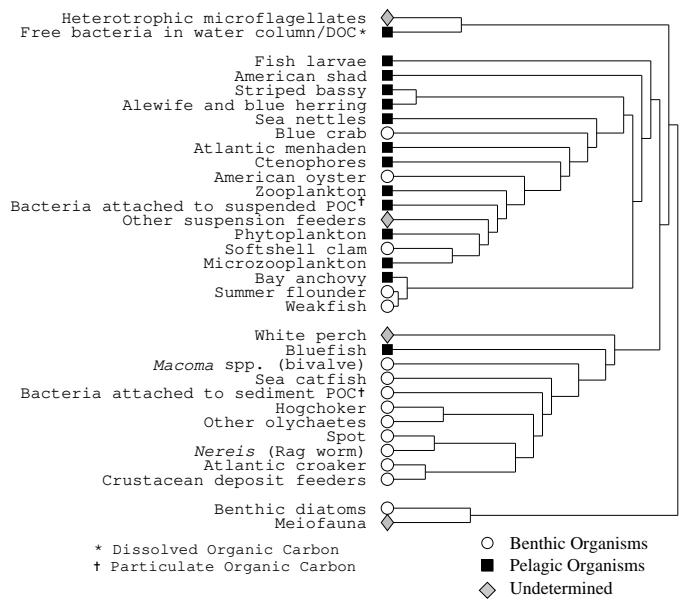}}
\end{center}
\caption{The dendrogram found by our method for Baird and Ulanowicz's food
web of marine organisms in the Chesapeake Bay~\cite{BU89}.}
\label{foodweb}
\end{figure}

Finally, for our third example, we take a network for which we do not have
any strong presuppositions about a ``correct'' division into communities.
This example is a true experiment to see what information the algorithm can
give us about a network whose structure is not wholly understood.  The
network in question is a food web, the web of trophic interactions (who
eats whom) of marine organisms living in the Chesapeake Bay.  The network
was assembled by Baird and Ulanowicz~\cite{BU89} and contains 33 vertices
representing the ecosystem's most prominent taxa.  The edges in a food web
are, technically, directed; they can be thought of as pointing from prey to
their predators, thus indicating the direction of energy (or carbon) flow
up the food chain.  Here however we have ignored the directed nature of the
network, considering the edges merely to be undirected indicators of
trophic interaction between taxon pairs.

The dendrogram produced for this food web by our community structure
algorithm is shown in Fig.~\ref{foodweb}.  As we can see, the algorithm
splits the network into two principle communities and a couple of smaller
peripheral ones.  We have shaded the vertices in the dendrogram to show
which taxa are surface dwellers in the bay (pelagic species) and which
bottom dwellers (benthic species).  A few species are of undetermined
status.  It is clear that our algorithm has in this case primarily
extracted from the network the distinction between pelagic and benthic
taxa.  Thus our results appear to imply that the food web in question can
be split roughly into separate surface- and bottom-dwelling subsystems,
with relatively weak interaction between the two.  A small number of
benthic species are found to belong more strongly to the pelagic community
than to the benthic one, perhaps indicating that a simple classification of
species by where they live is not telling the whole story for this system.
The results of our analysis might also be helpful in assigning a type to
the undetermined species in the network.

\section{Origins of community structure and assortative mixing}
\label{origins}
There is certainly more than one possible explanation for the presence of
community structure in a network, and different explanations may be
appropriate to different networks.  In the case of a social network, for
example, Jin~\etal~\cite{JGN01} have shown that communities can arise as a
result of the growth dynamics of a network.  If an acquaintance network
grows by the introduction of pairs of people to one another by a mutual
acquaintance, then an initial chance acquaintance with a member of a
certain community will lead to introductions to other members of that
community, so that one ultimately becomes linked to many of the community's
members and so becomes a member oneself.  Using a simple computer model of
this process, Jin~\etal\ found that even networks with no initial community
structure quickly develop such structure over time.  One can think of this
as a mechanism for the development of cliquishness in social networks.

This mechanism however is quite specific to social networks and could not
be easily applied, for example, to the food web studied in the last
section.  It also completely ignores any personal attributes of the actors
involved or affinities between actor pairs.  A more general and perhaps
more convincing explanation for community formation, which takes these
things into account, is that of assortative mixing,\footnote{The name
``assortative mixing'' comes from the epidemiology community, where this
effect has been studied extensively.  It is also sometimes called
``assortative matching'', particularly by ecologists.} which is the
tendency for nodes in a network to form connections preferentially to
others that are like them in some way.

\begin{table}[t]
\begin{center}
\begin{tabular}{l|r|cccc|c}
\multicolumn{2}{c|}{} & \multicolumn{4}{c|}{women}               \\
\cline{3-6}
\multicolumn{2}{c|}{} & black & hispanic & white & other & $a_i$ \\
\hline
\begin{rotate}{90}
\hbox{\hspace{-24pt}men}
\end{rotate}
& black    & 0.258 & 0.016 & 0.035 & 0.013 & 0.323 \\
& hispanic & 0.012 & 0.157 & 0.058 & 0.019 & 0.247 \\
& white    & 0.013 & 0.023 & 0.306 & 0.035 & 0.377 \\
& other    & 0.005 & 0.007 & 0.024 & 0.016 & 0.053 \\
\hline
\multicolumn{2}{r|}{$b_i$} 
           & 0.289 & 0.204 & 0.423 & 0.084 \\
\end{tabular}
\end{center}
\caption{The mixing matrix $e_{ij}$ and the values of $a_i$ and $b_i$ for
sexual partnerships in the San Francisco study described in the text.
After Morris~\cite{Morris95}.}
\label{sanfran}
\end{table}

An example of assortative mixing in social networks is mixing by race.
Table \ref{sanfran} shows data from the AIDS in Multiethnic Neighborhoods
study~\cite{CCKF92}, on mixing by race among sexual partners in the city of
San Francisco, California.  This part of the study focused on heterosexual
partnerships, and the rows and columns of the matrix represent men and
women in such partnerships, grouped by their (self-identified) race.
Diagonal elements of the matrix represent the fraction of survey
respondents in partnerships with members of their own group, and
off-diagonal those in partnerships with members of other groups.
Inspection of the figures shows that the matrix has considerably more
weight along its diagonal than off it, indicating that assortative mixing
does take place in this network.  One might well expect mixing of this type
to result in divisions within the community along lines of race, and we
will show shortly that, within the context of simulations of network
formation, assortative mixing can indeed give rise to community structure.

The amount of assortative mixing in a network can be characterized by
measuring how much of the weight in the mixing matrix falls on the
diagonal, and how much off it.  Let us define $e_{ij}$ to be the fraction
of all edges in a network that join a vertex of type~$i$ to a vertex of
type~$j$.  In the case of the matrix of Table~\ref{sanfran}, where the ends of
an edge always attach to one man and one woman, we also specify which index
corresponds to which type of end, which makes $e_{ij}$ asymmetric.  For
example, we could specify that the first index~$i$ represents the man and
the second~$j$ the woman.  For networks in which there is no corresponding
distinction, $e_{ij}$~will be symmetric.  The matrix should also satisfy
the sum rules
\begin{equation}
\sum_{ij} e_{ij} = 1,\qquad \sum_j e_{ij} = a_i,\qquad \sum_i e_{ij} = b_j,
\label{sumrules}
\end{equation}
where $a_i$ and $b_i$ are the fraction of each type of end of an edge that
is attached to vertices of type~$i$.  The values of $a_i$ and $b_i$ for the
San Francisco study are also shown in Table~\ref{sanfran}.  On graphs where
there is no distinction between the ends of edges, we will have $a_i=b_i$.

Now we can define a quantitative measure~$r$ of the level of assortative
mixing in the network thus~\cite{Newman02g}:
\begin{equation}
r = {\sum_i e_{ii} - \sum_i a_i b_i\over 1 - \sum_i a_i b_i}
  = {\Tr\ve - \norm{\ve^2}\over 1 - \norm{\ve^2}},
\label{defsr1}
\end{equation}
where $\ve$ is the matrix whose elements are the~$e_{ij}$, and the notation
$\norm{\vx}$ indicates the sum of the elements of the matrix~$\vx$.  We
call the quantity~$r$ the ``assortativity coefficient''.  It takes the
value~1 in a perfectly assortative network, since in that case the entire
weight of the matrix~$\ve$ lies along its diagonal and $\sum_i e_{ii}=1$.
Conversely, if there is no assortative mixing at all, then $e_{ij}=a_ib_j$
for all $i,j$ and $r=0$.  Networks can also be disassortative: vertices may
associate preferentially with others of different types---the ``opposites
attract'' phenomenon.  In that case, $r$~will take a negative value.

One can certainly imagine that assortative mixing might apply in other
types of networks as well.  For example, we saw in Section~\ref{examples}
that a food web of marine organisms apparently divided into communities
along lines of location---which species were surface dwellers (pelagic) and
which bottom dwellers (benthic).  It seems reasonable to hypothesize that
the evolution of new predatory relationships between species is biased by
the location of those species' living quarters, and hence that the network
structure would indeed reflect the pelagic/benthic division as a result of
assortative mixing by location.

We can test our hypothesis that assortative mixing could be responsible for
community formation in networks by computer simulation.  Given a mixing
matrix of the type shown in Table~\ref{sanfran}, we can create a random
network with the corresponding mixing pattern and any desired degree
distribution by the following algorithm.
\begin{enumerate}
\item First we choose degree distributions $p_k^{(i)}$ for each vertex
  type~$i$.  The quantity $p_k^{(i)}$ here denotes the probability that a
  randomly chosen vertex of type~$i$ will have degree~$k$.  We can also
  calculate the mean degree $z_i=\sum_k kp_k^{(i)}$ for each vertex type.
\item Next we choose a size for our graph in terms of the number $m$ of
  edges and draw $m$ edges from the desired distribution $e_{ij}$.  We
  count the number of ends of edges of each type~$i$, to give the
  sums~$m_i$ of the degrees of vertices in each class, and we calculate the
  expected number~$n_i$ of vertices of each type from $n_i=m_i/z_i$
  (rounded to the nearest integer).
\item We draw $n_i$ vertices from the desired degree
  distribution~$p_k^{(i)}$ for type~$i$.  Normally the degrees of these
  vertices will not sum exactly to~$m_i$ as we want them to, in which case
  we choose one vertex at random, discard it, and draw another from the
  distribution~$p_k^{(i)}$, repeating until the sum does equal~$m_i$.
\item We pair up the $m_i$ ends of edges of type~$i$ at random with the
  vertices we have generated, so that each vertex has the number of
  attached edges corresponding to its chosen degree.
\item We repeat from step~3 for each vertex type.
\end{enumerate}
\begin{figure}[t]
\begin{center}
\resizebox{8cm}{!}{\includegraphics{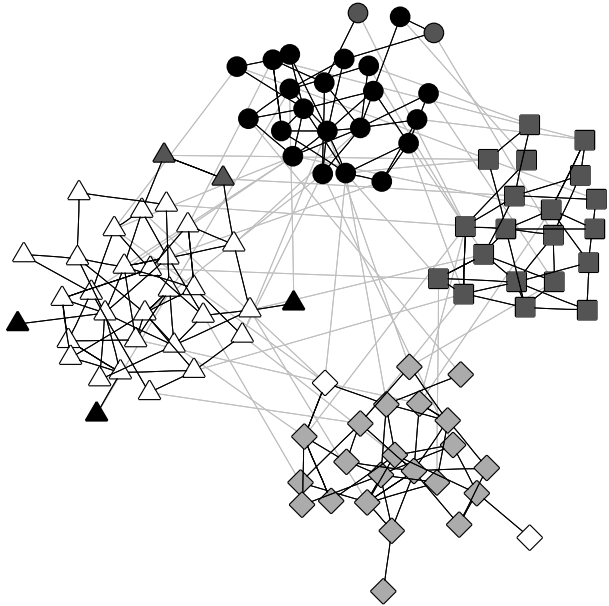}}
\end{center}
\caption{A network generated using the mixing matrix of Eq.~\eref{egmatrix}
and a Poisson degree distribution with mean $z=5$.  The four different
shades of vertices represent the four types, and the four shapes represent
the communities discovered by the community-finding algorithm of
Section~\ref{gn}.  The placement of the vertices has also been chosen to
accentuate the communites and show where the algorithm fails.  As we can
see, the correspondence between vertex type and the detected community
structure is very close; only nine of the 100 vertices are misclassified.}
\label{fourway}
\end{figure}
We have used this algorithm to generate example networks with desired
levels of assortative mixing.  For example, Fig.~\ref{fourway} shows an
undirected network of $n=100$ vertices of four different types, generated
using the symmetric mixing matrix
\begin{equation}
\ve = \begin{pmatrix}
        0.18 & 0.02 & 0.01 & 0.03 \\
        0.02 & 0.20 & 0.03 & 0.02 \\
        0.01 & 0.03 & 0.16 & 0.01 \\
        0.03 & 0.02 & 0.01 & 0.22 \\
      \end{pmatrix},
\label{egmatrix}
\end{equation}
which gives a value of $r=0.68$ for the assortativity coefficient.  A
simple Poisson degree distribution with mean~$z=5$ was used for all vertex
types.  The graph was then fed into the community finding algorithm of
Section~\ref{gn} and a cut through the resulting dendrogram performed at
the four-community level.  The communities found are shown by the four
shapes of vertices in the figure and correspond very closely to the real
vertex type designations, which are represented by the four different
vertex shades.  In other words, by introducing assortative mixing by vertex
type into this network, we have created vertex-type communities that
register in our community finding algorithm in exactly the same way as
communities in naturally occurring networks.  This strongly suggests that
assortative mixing could indeed be an explanation for the occurrence of
such communities, although it is worth repeating that other explanations
are also possible.

\section{Other types of assortative mixing}
\begin{figure}[t]
\begin{center}
\resizebox{8cm}{!}{\includegraphics{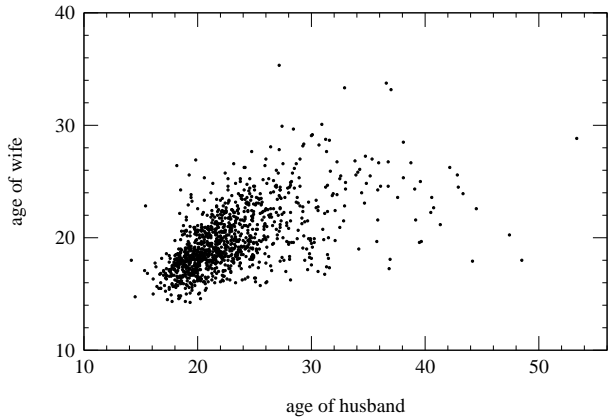}}
\end{center}
\caption{Scatter plot of the ages at first marriage of 1141 women
interviewed in the 1995 National Survey of Family Growth, and their
spouses.}
\label{marriage}
\end{figure}

Assortative mixing can depend on vertex properties other than the simple
enumerative properties discussed in the preceding section.  For example, we
can also have assortative mixing by scalar characteristics, either discrete
or continuous.  A classic example of such mixing, much studied in the
sociological literature, is acquaintance matching by age.  In many
contexts, people appear to prefer to associate with others of approximately
the same age as themselves.  For example, consider Fig.~\ref{marriage},
which shows the ages at marriage of the male and female members of 1141
married couples drawn from the US National Survey of Family
Growth~\cite{NSFG97}.  Each point in the figure represents one couple, its
position along the horizontal and vertical axes corresponding to the ages
of the husband and wife respectively.  The study was based on interviews
with women, and was limited to those of childbearing age, so the vertical
axis cuts off around~40.  Also only the first marriage for each woman
interviewed is shown, even if she married more than once.  Despite these
biases however, the figure reveals a clear trend: people prefer to marry
others of an age close to their own.

It is perhaps stretching a point a little to consider first marriage ties
between couples as forming a social network, since people have at most one
first marriage and hence would have a maximum degree of one within the
network.  Here, however, we consider marriage age as a proxy for the ages
of sexual partners in general, and conjecture that a similar age preference
will be seen in non-married partners also.  A recent study by
Garfinkel~\etal~\cite{GGM02} appears to support this conjecture.

Assortative mixing according to scalar characteristics can result in the
formation of communities, just as in the case of discrete characteristics.
One could have separate communities formed of old and young people, for
instance.  However, it is also possible that we do not get well-defined
communities, but instead get an overlapping set of groups with no clear
boundaries, ranging for example from low age to high age.  In the
sociological literature such a continuous gradation of one community into
another is called ``stratification'' of the network.

As with assortative mixing on discrete characteristics, one can define an
assortativity coefficient to quantify the extent to which mixing is biased
according to scalar vertex properties.  To do this, we define $e_{xy}$ to
be the fraction of edges in our network that connect a vertex of
property~$x$ to another of property~$y$.  The matrix $e_{xy}$ must satisfy
sum rules as before, of the form
\begin{equation}
\sum_{xy} e_{xy} = 1,\qquad \sum_y e_{xy} = a_x,\qquad \sum_x e_{xy} = b_y,
\label{sumrulesxy}
\end{equation}
where $a_x$ and $b_y$ are, respectively, the fraction of edges that start
and end at vertices with ages~$x$ and~$y$.  Then the appropriate definition
for the assortativity coefficient is
\begin{equation}
r = {\sum_{xy} xy(e_{xy}-a_xb_y)\over\sigma_a\sigma_b},
\label{rscalar}
\end{equation}
where $\sigma_a$ and $\sigma_b$ are the standard deviations of the
distributions $a_x$ and $b_y$.  The reader will no doubt recognize this
definition of $r$ as the standard Pearson correlation coefficient for the
quantities $x$ and~$y$.  It takes values in the range $-1\le r\le1$ with
$r=1$ indicating perfect assortative mixing, $r=0$ indicating no
correlation between $x$ and~$y$, and $r=-1$ indicating perfect
disassortative mixing, i.e.,~perfect anticorrelation between $x$ and~$y$.

If we take the marriage data from Fig.~\ref{marriage}, for example, and
feed it into Eq.~\eref{rscalar}, we find that $r=0.57$, indicating once
again that mixing is strongly assortative (as is in any case obvious from
the figure).

Mixing could also depend on vector or even tensor characteristics of
vertices.  One example would be mixing by geographical location, which
could be regarded as a two-vector.  It seems highly likely that if one were
to record both acquaintance patterns and geographical location for actors
in a social network, one would discover that acquaintance is strongly
dependent on geography, with people being more likely to know others who
live in the same part of the world as themselves.

\subsection{Mixing by vertex degree}
\label{degree}
We will spend the rest of this article examining one particular case of
mixing according to a scalar vertex property, that of mixing by vertex
degree, which has been studied for some time in the social networks
literature and has recently attracted attention in the mathematical and
physical literature also.  Krapivsky and Redner~\cite{KR01} for instance
found in studies of the preferential attachment model of Barab\'asi and
Albert~\cite{BA99b} that edges did not fall between vertices independent of
their degrees.  Instead there was a higher probability to find some degree
combinations at the ends of edges than others.
Pastor-Satorras~\etal~\cite{PVV01} showed for data on the structure of the
Internet at the level of autonomous systems that the degrees of adjacent
vertices were anticorrelated, i.e.,~that high-degree vertices prefer to
attach to low-degree vertices, rather than other high-degree ones---the
network is disassortative by degree.  To demonstrate this, they measured
the mean degree degree $\av{k_{\rm nn}}$ of the nearest-neighbours of a
vertex, as a function of that vertex's degree~$k$.  They found that
$\av{k_{\rm nn}}$ decreases with increasing~$k$, approximately
as~$k^{-1/2}$.  That is, the mean degree of your neighbours goes down as
yours goes up.  Maslov and Sneppen~\cite{MS02b} have offered an explanation
of this result in terms of ensembles of graphs in which double edges
between vertices are forbidden.  Maslov and Sneppen also showed in a
separate paper~\cite{MS02a} that the protein interaction network of the
yeast \textit{S. Cerevisiae} displays a similar sort of disassortative
mixing.

An alternative way to quantify assortative mixing by degree in a network is
to use an assortativity coefficient of the type described in the previous
section~\cite{Newman02f}.  Let us define $e_{jk}$ to be the fraction of
edges in a network that connect a vertex of degree~$j$ to a vertex of
degree~$k$.  (As before, if the ends of an edge connect different types of
vertices, then the matrix will be asymmetric, otherwise it will be
symmetric.)  In fact, we define $j$ and $k$ to be the ``excess degrees'' of
the two vertices, i.e.,~the number of edges incident on them less the one
edge that we are looking at at present.  In other words, $j$~and $k$ are
one less than the total degrees of the two vertices.  This designation
turns out to be mathematically convenient for many developments.  If the
degree distribution of the network as a whole is~$p_k$, then the
distribution of the excess degree of the vertex at the end of a randomly
chosen edge is
\begin{equation}
q_k = {(k+1)p_{k+1}\over z},
\end{equation}
where $z=\sum_k kp_k$ is the mean degree~\cite{Newman02d}.  Then one can
define the assortativity coefficient to be
\begin{equation}
r = {\sum_{jk} jk(e_{jk}-q_jq_k)\over\sigma_q^2},
\end{equation}
where $\sigma_q$ is the standard deviation of the distribution~$q_k$.  On a
directed or similar network, where the ends of an edge are not the same and
$e_{jk}$ is asymmetric, this generalizes to
\begin{equation}
r = {\sum_{jk} jk(e_{jk}-q^a_jq^b_k)\over\sigma_a\sigma_b},
\label{rdir}
\end{equation}
where $\sigma_a$ and $\sigma_b$ are the standard deviations of the
distributions $q^a_k$ and $q^b_k$ for the two types of ends.  (The measure
introduced by Pastor-Satorras~\etal~\cite{PVV01} can also be expressed
simply in terms of the matrix~$e_{jk}$: it is $\av{k_{\rm nn}} =
\sum_j je_{jk}/q_k$.  Maslov and Sneppen~\cite{MS02a,MS02b} gave entire plots
of the raw~$e_{jk}$, using colours to code for different values.  These
plots are however rather difficult to interpret by eye.)

\begin{table}[t]
\begin{center}
\begin{tabular}{r|l|c|r|c|c}
& network                    & type       & size $n$      & assortativity $r$ & ref. \\
\hline
\begin{rotate}{90}
\hbox{\hspace{-3.9em}social}
\end{rotate}
& physics coauthorship       & undirected & $52\,909$     & $\phm0.363$       & a    \\
& biology coauthorship       & undirected & $1\,520\,251$ & $\phm0.127$       & a    \\
& mathematics coauthorship   & undirected & $253\,339$    & $\phm0.120$       & b    \\
& film actor collaborations  & undirected & $449\,913$    & $\phm0.208$       & c    \\
& company directors          & undirected & $7\,673$      & $\phm0.276$       & d    \\
& email address books        & directed   & $16\,881$     & $\phm0.092$       & e    \\
\hline
\begin{rotate}{90}
\hbox{\hspace{-2.6em}technol.}
\end{rotate}
& Internet                   & undirected & $10\,697$     & $-0.189$          & f    \\
& World-Wide Web             & directed   & $269\,504$    & $-0.067$          & g    \\
& software dependencies      & directed   & $3\,162$      & $-0.016$          & h    \\
\hline
\begin{rotate}{90}
\hbox{\hspace{-4.1em}biological}
\end{rotate}
& protein interactions       & undirected & $2\,115$      & $-0.156$          & i    \\
& metabolic network          & undirected & $765$         & $-0.240$          & j    \\
& neural network             & directed   & $307$         & $-0.226$          & k    \\
& marine food web            & directed   & $134$         & $-0.263$          & l    \\
& freshwater food web        & directed   & $92$          & $-0.326$          & m    \\
\end{tabular}
\end{center}
\caption{Size~$n$ and degree assortativity coefficient~$r$ for a number
real-world networks.  Social networks: coauthorship networks of
(a)~physicists and biologists~\cite{Newman01a} and
(b)~mathematicians~\cite{GI95}; (c)~collaborations (co-starring
relationships) of film actors~\cite{WS98,NSW01}; (d)~directors of Fortune
1000 companies for 1999, in which two directors are connected if they sit
on the board of directors of the same company~\cite{DYB01,NSW01};
(e)~network of email address books of computer users~\cite{NFB02}.
Technological networks: (f)~network of direct peering relationships between
autonomous systems on the Internet, April~2001~\cite{Chen02}; (g)~network
of hyperlinks between pages in the World-Wide Web domain
\texttt{nd.edu} \textit{circa} 1999~\cite{AJB99}; (h)~network of
dependencies between software packages in the GNU/Linux operating
system~\cite{Newman02g}.  Biological networks: (i)~protein--protein
interaction network in the yeast \textit{S. Cerevisiae}~\cite{Jeong01};
(j)~metabolic network of the bacterium \textit{E. Coli}~\cite{Jeong00};
(k)~neural network of the nematode worm
\textit{C.~Elegans}~\cite{WSTB86,WS98}; tropic interactions between species
in the food webs of (l)~Ythan Estuary, Scotland~\cite{HBR96} and (m)~Little
Rock Lake, Wisconsin~\cite{Martinez91}.}
\label{rtab}
\end{table}

In Table~\ref{rtab} we show values of $r$ measured for a variety of
different real-world networks.  The networks shown are divided into social,
technological and biological networks, and a particularly striking feature
of the table is that the values of $r$ for the social networks are all
positive, indicating assortative mixing by degree, while those for the
technological and biological networks are all negative, indicating
disassortative mixing.  It is not clear at present why this should be,
although explanations for the observed mixing behaviours have been proposed
in some specific cases~\cite{MS02b,Newman02g}.

As with the mixing by discrete enumerative characteristics discussed in
Section~\ref{origins}, we can also investigate the effects of assortative
mixing by looking at computer generated networks with particular types of
mixing.  Unfortunately, no simple algorithm exists for generating graphs
mixed by vertex degree analogous to that of Section~\ref{origins} (see
Refs.~\cite{DMS02} and~\cite{Newman02g}) and one is forced to resort to
Monte Carlo generation of graphs using Metropolis--Hastings type algorithms
of the sort widely used for graph generation in mathematics and
quantitative sociology.  Such algorithms however are straightforward to
implement.  For the present case, we take the simple example form
\begin{equation}
e_{jk} = \cN \e^{-(j+k)/\kappa} \biggl[ {j+k\choose j} p^j q^k +
                                   {j+k\choose k} p^k q^j \biggr],
\label{binomial}
\end{equation}
where $p+q=1$, $\kappa>0$, and $\cN=\half(1-\e^{-1/\kappa})$ is a
normalizing constant.  This means that the distribution of the sum $j+k$ of
the excess degrees at the ends of an edge falls off as a simple
exponential, while that sum is distributed between the two ends binomially,
the parameter~$p$ controlling the assortative mixing.  For values of $p$
ranging from 0 to $\half$ we get various values of the assortativity~$r$,
both positive and negative, passing through zero at
$p_0=\half-\frac14\sqrt{2}=0.1464\ldots$

\begin{figure}[t]
\begin{center}
\resizebox{\textwidth}{!}{\includegraphics{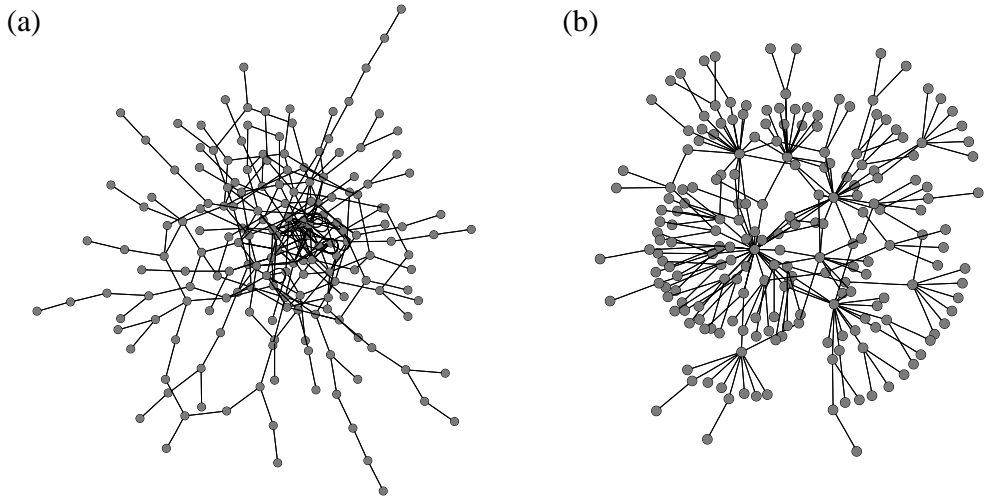}}
\end{center}
\caption{The giant component of two graphs generated using a Monte Carlo
procedure with edge distribution given by Eq.~\eref{binomial} with
$\kappa=10$ and (a)~$p=0.5$ and (b)~$p=0.05$.}
\label{graphs}
\end{figure}

As an example, we show in Fig.~\ref{graphs} the giant components of two
graphs of this type generated using the Monte Carlo method.  One of them,
graph~(a), is assortatively mixed by degree, while the other, graph~(b), is
disassortatively mixed.  The difference between the two is clear to the
eye.  In the first case, because the high degree vertices prefer to attach
to one another, there is a central ``core'' to the network, composed of
these high-degree vertices, and a straggling periphery of low-degree
vertices around it.  In epidemiology a dense central portion of this type
is called a ``core group'' and is thought to be capable of acting as a
reservoir for disease, keeping diseases circulating even when the density
of the network as a whole is too low to maintain endemic infection.  In
social network analysis one also talks of ``core/periphery'' distinctions
in networks, another concept that mirrors what we see here.  In the second
graph, which is disassortative, a contrasting picture is evident: the
high-degree vertices prefer not to associate with one another, and are as a
result scattered widely over the network, producing a more uniform
appearance.

\begin{figure}[t]
\begin{center}
\resizebox{8cm}{!}{\includegraphics{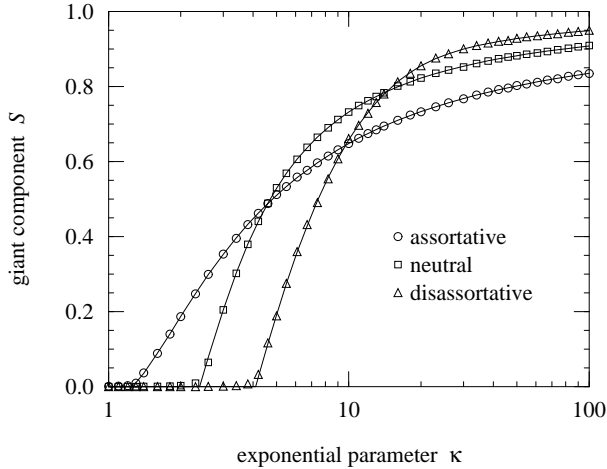}}
\end{center}
\caption{The size of the giant component as a function of graph size for
graphs with the edge distribution given in Eq.~\eref{binomial}, for three
different values of the parameter~$p$, which controls the assortativity.
The points are simulation results for graphs of $N=100\,000$ vertices while
the solid lines are the analytic solution for the same quantity given by
Newman~\cite{Newman02f}.  Each point is an average over ten graphs; the
resulting statistical errors are smaller than the symbols.  The values
of~$p$ are 0.5 (circles), $p_0=0.146\ldots$ (squares), and $0.05$
(triangles).}
\label{gc}
\end{figure}

To shed more light on the effects of assortativity, we show in
Fig.~\ref{gc} the size of the largest component in networks of this type as
the degree distribution parameter $\kappa$ is varied, for various values
of~$p$.  For low values of $\kappa$ the mean degree of the network is
small, and the resulting density of edges is too low to produce percolation
in the network, so there is no giant component.  As $\kappa$ increases,
however, there comes a point, clearly visible on the plot, at which the
edge density is great enough to form a giant component.  Figure~\ref{gc}
reveals two interesting features of this transition.  First, the position
of the transition, the value of the parameter $\kappa$ at which it takes
place, is smaller in assortatively mixed networks than in disassortative
ones.  In other words, it appears that the presence of assortativity in the
degree correlation pattern allows the network to percolate more easily.
This result is intuitively reasonable: the core group of the assortative
network seen in Fig.~\ref{graphs}a has a higher density of edges than the
network as a whole and so one would expect percolation to take place in
this region before it would in a network with the same average density but
no core group.

Second, the figure shows that, even though the assortative network
percolates more easily than its disassortative counterpart, its largest
component does not grow as large as that of the disassortative network in
the limit where $\kappa$ becomes large.  This too can be understood in
simple terms: percolation occurs more easily when there is a core group,
but is also largely confined to that core group and so does not spread to
as large a portion of the network as it would in other cases.

In epidemiological terms, one could think of these two results as
indicating that assortative networks will support the spread and
persistence of a disease more easily than disassortative ones, because they
possess a core group of connected high-degree vertices.  But the disease is
also restricted mostly to that core group.  In a disassortative network,
although percolation and hence epidemic disease requires a denser network
to begin with, when it does happen it will affect a larger fraction of the
network, because it is not restricted to a core group.

\section{Conclusions}
\label{concs}
In this article we have examined two related properties of networks:
community structure and assortative mixing.  We have described a new
algorithm for finding groups of tightly-knit vertices within
networks---communities in our nomenclature---which is based on the
calculation of an ``edge betweenness'' index for network edges.  The
algorithm appears to be successful at detecting known community structure
in various example networks, and we have found that a number of real-world
networks do indeed possess community structure to a greater or lesser
degree.

Turning to possible explanations for this structure we have suggested that
assortative mixing, the preferential association of vertices in a network
with others that are like them in some way, is one possible mechanism for
community formation.  We have defined a measure of the strength of
assortative mixing and applied it, for example, to data on mixing by race
in social networks, showing that there is strong assortativity in this
case, at least for the survey data that we have examined.  We have also
given a simple algorithm for creating networks with assortative mixing
according to discrete characteristics imposed upon the vertices, and used
it to generate example networks which, when fed into our community
detection algorithm, reveal strong community structure similar to that seen
in the real-world data.  This lends some conviction to the theory that
assortative mixing could, at least in some cases, be a contributing factor
in the formation of communities within networks.

We have also looked at assortative mixing by scalar characteristics of
vertices, such as the age of individuals in a social network, and
particularly vertex degree.  By measuring mixing of the latter type for a
variety of different networks, we have shown that social networks appear
often to be assortatively mixed by degree, while technological and
biological networks appear normally to be disassortative.  Using computer
generated model networks we have also shown that assortativity by vertex
degree makes networks percolate more easily---they develop a giant
component for a lower average edge density than a similar network with
neutral or disassortative mixing.  Conversely, however, disassortative
networks tend to have larger giant components when they do develop.  These
findings have implications for epidemiology, for example: they imply that a
disease spreading on a network that is assortatively mixed, as most social
networks appear to be, would reach epidemic proportions more easily than on
a disassortative network, but that the epidemic might ultimately affect
fewer people than in the disassortative case.

Looking ahead, some obvious next steps in the studies presented here are
the application of community finding algorithms to other networks, the
study of mixing patterns in other networks, and theoretical investigations
of the effects of assortative mixing and other network correlations on
network structure and function, including for instance network resilience
and network epidemiology.  A number of authors have already started work on
these problems~\cite{HHJ02,WH03,BPV02,VM02,Newman02f,Newman02g}.

\section*{Acknowledgements}
The authors would like to thank Jennifer Dunne, Neo Martinez and Doug White
for help assembling and interpreting the data used in Figs.~\ref{zachary}
and~\ref{foodweb}, and L\'aszl\'o Barab\'asi, Jerry Davis, Jennifer Dunne,
Jerry Grossman, Hawoong Jeong, Neo Martinez and Duncan Watts for providing
data used in the calculations for Table~\ref{rtab}.  The marriage data for
Fig.~\ref{marriage} were provided by the Inter-University Consortium for
Political and Social Research at the University of Michigan.  This work was
supported in part by the National Science Foundation under grants
DMS--0109086 and DMS--0234188.

\end{document}